\documentclass[11pt]{article}

\usepackage[final]{acl}

\usepackage{times}
\usepackage{latexsym}

\usepackage[T1]{fontenc}

\usepackage[utf8]{inputenc}

\usepackage{microtype}

\usepackage{inconsolata}

\usepackage{graphicx}

\usepackage[ruled,vlined]{algorithm2e}
\usepackage{booktabs}
\usepackage{amsmath}
\usepackage{multirow}
\usepackage{float}
\usepackage{makecell}
\usepackage{graphicx}
\usepackage[misc]{ifsym}
\usepackage{enumitem}
\usepackage{caption}
\usepackage{amsfonts}
\usepackage{utfsym}
\usepackage{comment}

\usepackage{amssymb}
\usepackage{bbding}
\usepackage{wasysym}

\usepackage{booktabs}
\usepackage{tabularx}

\title{ControlAudio: Tackling Text-Guided, Timing-Indicated and Intelligible Audio Generation via Progressive Diffusion Modeling}

\author{%
  Yuxuan Jiang$^{1,2,}\thanks{Equal Contribution.}$, 
  Zehua Chen$^{1,2,}\footnotemark[1]^{,}\thanks{Corresponding Authors: Zehua Chen and Jun Zhu.}$, 
  Zeqian Ju$^{3}$, 
  Yusheng Dai$^{2,4}$, 
  Weibei Dou$^{1}$, 
  Jun Zhu$^{1,2,}\footnotemark[2]$ \\ 
  $^{1}$Tsinghua University  $^{2}$Shengshu AI \\  
  $^{3}$University of Science and Technology of China $^{4}$Monash University
}

\begin{document}
\maketitle
\begin{abstract}
Text-to-audio (TTA) generation with fine-grained control signals, \textit{e.g.}, precise timing control or intelligible speech content, has been explored in recent works. However, constrained by data scarcity, their generation performance at scale is still compromised. In this study, we recast controllable TTA generation as a multi-task learning problem and introduce a progressive diffusion modeling approach, ControlAudio. Our method adeptly fits distributions conditioned on more fine-grained information, including text, timing, and phoneme features, through a step-by-step strategy. First, we propose a data construction method spanning both annotation and simulation, augmenting condition information in the sequence of text, timing, and phoneme. Second, at the model training stage, we pretrain a diffusion transformer (DiT) on large-scale text-audio pairs, achieving scalable TTA generation, and then incrementally integrate the timing and phoneme features with unified semantic representations, expanding controllability. Finally, at the inference stage, we propose progressively guided generation, which sequentially emphasizes more fine-grained information, aligning inherently with the coarse-to-fine sampling nature of DiT. Extensive experiments show that ControlAudio achieves state-of-the-art performance in terms of temporal accuracy and speech clarity, significantly outperforming existing methods on both objective and subjective evaluations. Demo samples are available at:~\url{https://control-audio.github.io/Control-Audio/}.

\end{abstract}

\section{Introduction}

Text-to-audio (TTA) generation systems aim at synthesizing high-fidelity audio samples that are consistent with the given natural language description,~\textit{e.g.}, "A bird is chirping"~\cite{liu2023audioldm,ghosal2023text,huang2023make,evans2024fast}.  
Recent efforts are exploring more fine-grained control for TTA systems, which can be categorized into two main classifications.
The first group adds precise timing control,~\textit{e.g.}, "A bird is chirping, at 2-5 seconds", with innovations spanning conditioning techniques~\cite{wang2025audiocomposer,xie2024picoaudio} and training-free latent manipulation~\cite{jiang2025freeaudio}.
The second group works on intelligible audio generation,~\textit{e.g.}, "A bird is chirping, and a man is saying: 'it's a very sunny day'", by introducing additional modules to encode both audio and speech semantic information~\cite{lee2024voiceldm,jung2025voicedit}.
However, as expensive to collect large-scale text-audio datasets with precise timing and speech information, their controllable generation performance at scale remains limited, and none of the prior work explores~\textit{timing-controlled and intelligible TTA generation},~\textit{e.g.}, "A bird is chirping, at 0-5 seconds, and then a man is saying: 'it's a very sunny day', at 7-10 seconds", within a unified framework.

In this work, we propose ControlAudio, a progressive diffusion modeling approach to progressively capture the target distribution conditioned on fine-grained information, $ (\text{text, timing, phoneme})$, enabling controllable TTA generation at scale.
Our designs cover data construction and representation, model training, as well as guided sampling, each of which progressively integrates more fine-grained condition information, thereby expanding controllability at scale. 
In data construction, we collect large-scale $ \langle \text{text, audio} \rangle $ pairs, and then construct more expensive datasets, $ \langle \text{text, timing, audio} \rangle $ and $ \langle \text{text, timing, phoneme, audio} \rangle $, with both annotation and simulation methods, predefining the target distribution of each training stage.
For the representation of text and timing information, we develop a structural prompt, enabling a pre-trained text encoder to precisely encode them without fine-tuning.
Given the timing indication, namely the duration of the speech event, we naturally extend the vocabulary of the same encoder with phoneme tokens, realizing unified semantic modeling for text, timing, and phoneme features with a single text encoder.

With target distributions predefined by datasets constructed above, we introduce progressive diffusion training, fulfilling high-quality TTA synthesis at pre-training and gradually integrating fine-grained control signals at continual learning stages.
At the first stage, we pre-train a diffusion transformer (DiT) in the latent space directly compressed from the waveform space, solely conditioned on text indication, achieving high-fidelity TTA at scale.
At the second stage, we fine-tune the latent DiT on both text and timing conditions, enabling the model to precisely control the timing windows of each sound event.
In controllable TTA generation, a common issue is the sacrifice of text-conditioned synthesis quality without fine-grained conditions~\cite{wang2025audiocomposer}.
Hence, in ControlAudio, we switch the condition between the text condition and the $ \langle \text{text, timing} \rangle $ condition at the second stage, avoiding catastrophic forgetting in progressive training. 
At the final stage, given the audio generation prior learned in prior stages, we continually train the diffusion model by switching the condition among text, $ \langle \text{text, timing} \rangle $, and $ \langle \text{text, timing, phoneme} \rangle $, achieving high-fidelity audio synthesis conditioned on flexible indication.

In generation, diffusion models demonstrate a coarse-to-fine sampling nature. Along the entire trajectory, they generate large-scale features at the early stage and synthesize fine-grained details in the following steps, iteratively refining the generation results. 
In controllable TTA systems, condition signals show diverse control granularity as well.
Hence, for timing-controlled and intelligible audio generation, we design progressive guided sampling, where the timing condition first guides the sampling to indicate the timing windows as large-scale features and then the phoneme condition is introduced to indicate the speech content as small-scale features.
In comparison with a fixed guidance signal, our method gradually emphasizes more fine-grained condition information, inherently aligned with the diffusion sampling process.
Extensive experiments demonstrate that ControlAudio achieves state-of-the-art performance on controllable audio generation tasks, significantly outperforming existing methods in both objective and subjective evaluations of temporal accuracy and speech clarity.

\begin{figure*}[t]
\centerline{\includegraphics[width=17.1cm]{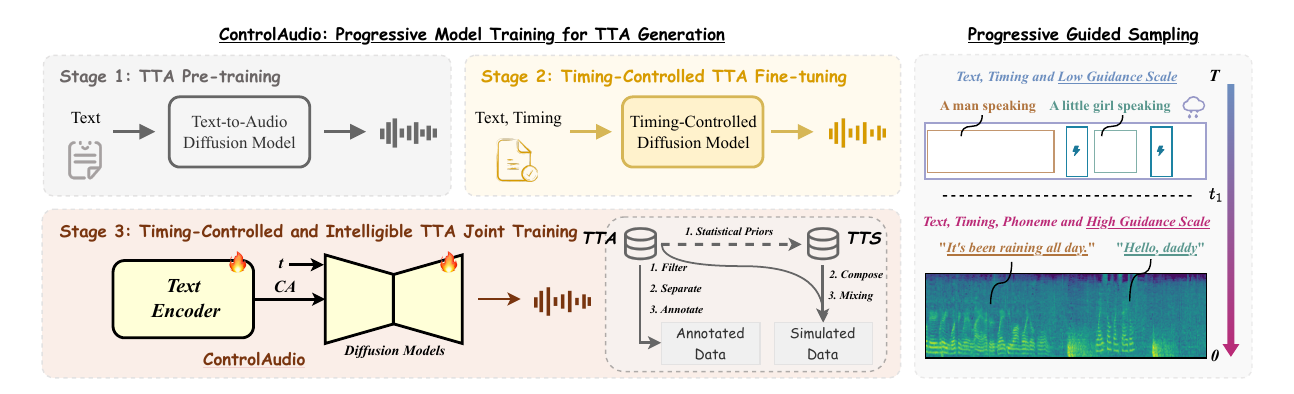}}
\caption{The end-to-end Progressive Diffusion Modeling of ControlAudio, which combines a progressive model training with a progressive guided sampling process for decoupled control of temporal structure and speech content.}
\label{fig:model}
\vspace{-0.4mm}
\end{figure*}

\section{Related Work}

\subsection{Controllable TTA Generation}

Recent works have explored temporal control in TTA generation through two main paradigms. Training-based methods, such as MC-Diffusion~\cite{guo2024audio}, PicoAudio~\cite{xie2024picoaudio} and Audio ControlNet~\cite{zhu2026audio}, rely on predefined event classes, which limits their flexibility for open-domain prompts. More recent approaches, including AudioComposer~\cite{wang2025audiocomposer}, PicoAudio2~\cite{zheng2025picoaudio2} and DegDiT~\cite{liu2026degdit}, adopt natural language as the control interface, but often face ambiguity when describing complex multi-event scenarios. 
Alternatively, training-free methods such as TG-Diff~\cite{du2024controllable} and FreeAudio~\cite{jiang2025freeaudio} enforce temporal alignment during inference, but typically incur high computational overhead and struggle in dense event settings. 
In parallel, generating intelligible speech remains a fundamental challenge. Most TTA models produce speech as indistinct vocalizations. While models like VoiceLDM~\cite{lee2024voiceldm}, VoiceDiT~\cite{jung2025voicedit} and UmbraTTS~\cite{glazer2025umbratts} achieve high-quality speech synthesis, they are designed as specialized TTS systems and lack unified control over general audio events. Moreover, prior work on controllable dialogue, such as CoVoMix2~\cite{zhang2025covomix2}, primarily focuses on speech-only scenarios.
To address these limitations, we propose a unified framework for timing-controlled generation of general audio events and intelligible multi-speaker dialogue.

\subsection{Progressive Modeling}

In recent cross-modal generation tasks, such as video or avatar generation conditioned on diverse control signals~\cite{zhao2025controlvideo,zhu2026causal,OmniHuman,hunyuancustom}, progressive modeling has proven effective in handling multi-condition video generation. 
However, its advantages have only been partially explored in TTA generation, such as in long-form narrative audio~\cite{guo2025audiostory} and video-to-audio~\cite{tian2025audiox,dai2026omni2sound} generation, and have not yet been extended to controllable TTA generation, where precise timing control and intelligible speech remain critical yet largely unresolved challenges. To address this gap, ControlAudio leverages progressive modeling to enable stage-wise controllability and fine-grained regulation across different levels.

\section{Preliminary}

\subsection{Diffusion-based TTA Generation}

Diffusion-based~\cite{peebles2023scalable,li2024qa} TTA models are typically trained to learn a conditional reverse of a data-to-noise forward process~\cite{ho2020denoising}, progressively removing noise from an initial random state conditioned on a text prompt over multiple diffusion steps. This framework consists of three main modules: 1) an audio varational autoencoder (VAE), responsible for transforming the audio sample into a compressed latent representation while ensuring the reconstruction quality; 2) a pretrained text encoder, which encodes a text prompt into conditioning embeddings; and 3) a latent diffusion model, which predicts the denoised audio latents conditioned on the text embeddings. 
In ControlAudio, we employ a DiT-based architecture to ensure scalability~\cite{evans2025stable}, conditioned on the text, timing, and phoneme embeddings to generate the latent audio representation directly compressed from the waveform, without cascaded decoding~\cite{liu2023audioldm,guo2024audio,dai2025latent}. 

\subsection{Classifier-Free Guidance}

Classifier-Free Guidance (CFG) \cite{ho2022classifier,wang2025audiomog} emphasizes the guidance of a conditioning signal $c$ during sampling. At each sampling step, CFG-guided diffusion models produce two predictions: a conditional estimation $\epsilon_{\theta}(x_t, c)$ and an unconditional estimation $\epsilon_{\theta}(x_t, \emptyset)$. Then the final prediction is obtained by extrapolating these two terms with a guidance scale $w > 1$:
\begin{equation}
\hat{\epsilon}_{\theta}(x_t, c) = \epsilon_{\theta}(x_t, \emptyset) + w \cdot \big(\epsilon_{\theta}(x_t, c) - \epsilon_{\theta}(x_t, \emptyset)\big).
\end{equation}
Typically, a larger guidance scale $w$ encourages stronger alignment with the condition, which may increase fidelity while sacrificing diversity.

\section{ControlAudio}

\subsection{Motivation}

As discussed above, current TTA generation quality has been advanced with latent diffusion models, while the quality of controllable generation,~\textit{e.g.}, precise timing control or intelligible speech control, is still limited. 
Although diverse innovations have been proposed, their synthesis quality at scale is still compromised by data scarcity.
Moreover, previous research rarely achieves versatile TTA generation, namely, integrating additional fine-grained control signals while preserving high-fidelity audio generation solely conditioned on text.

In this work, we propose a progressive diffusion modeling design covering data construction and representation, model training, and guided sampling to tackle these difficulties, achieving text-guided, timing-indicated, and intelligible audio generation with a single diffusion model. 
Figure~\ref{fig:model} illustrates our overall progressive strategy.

\subsection{Dataset Construction} 

\noindent\textbf{Data Scarcity.} 
For TTA generation, we can collect various publicly available datasets, which comprise millions of weakly-labeled text-audio pairs~(Appendix~\ref{sec:pre_data}), supporting high-quality synthesis at scale.
However, these datasets typically contain only high-level textual descriptions, lacking the fine-grained annotations required for controllable synthesis~\cite{wang2025audioatlas}. Specifically, training timing-controlled and intelligible TTA generation requires datasets that combine speech with general audio events under precise timing annotations. Yet, such datasets are rare: existing timing-annotated audio datasets are limited in scale and lack transcriptions for speech segments, while publicly available speech datasets do not have reliable temporal labels. To overcome this limitation, we first construct a multi-source dataset.

\noindent\textbf{Annotated Data.} Our data annotation pipeline begins with the AudioSet-SL~\cite{hershey2021benefit} dataset, chosen for its reliable temporal annotations while lacking corresponding speech transcripts. To create the ControlAudio dataset, we first select all clips containing "human speech" and then extract a clean speech track from each using a dual-demixing strategy inspired by MTV~\cite{weng2025audio} that leverages both MVSEP~\cite{ZFTurbo23} and Spleeter~\cite{hennequin2020spleeter}. The clean track is subsequently segmented into individual events using the original timestamps. 
Finally, each segmented event is transcribed using Gemini 2.5 Pro\footnote{\url{https://deepmind.google/models/gemini/pro/}}. 
Further details of the pipeline and prompt design are provided in Appendices~\ref{sec:anno_data} and~\ref{sec:alm}.
This transcription process enriches the dataset~\textit{i.e.}, expanding condition to $ \langle \text{text, timing, phoneme} \rangle $ for fine-grained control. For example, a generic annotation like (man speaking, <3.00,5.00>) is transformed into a specific, content-rich event~(man speaking: "It's been raining all day.", <3.00,5.00>). 

\noindent\textbf{Simulated Data.} To further expand our dataset, we construct a large-scale simulated dataset guided by real-world data distribution. Specifically, we first analyze the AudioSet-SL dataset to derive statistical priors on speech activity patterns, with further details provided in Appendix~\ref{app:simu_data}. These distributions guide our synthesis process, which proportionally simulates two main scenarios: single-speaker scenarios (\textit{monologue}) are created by combining multiple utterances from the same speaker in LibriTTS-R, while multi-speaker scenarios (\textit{dialogue}) are formed by sampling from different speakers. After composing the speech samples, we simulate a plausible temporal arrangement for the utterances. Finally, the composed speech is mixed with non-speech backgrounds from WavCaps~\cite{mei2024wavcaps} and VGG-Sound~\cite{chen2020vggsound} at a signal-to-noise ratio sampled from a uniform 2 to 10~dB range~\cite{jung2025voicedit}. Through this simulation pipeline, we generate an additional 171,246 complex audio scenes, significantly expanding the scale and diversity of our training data.

\begin{figure}[t]
\centering
\centerline{\includegraphics[width=7.5cm]{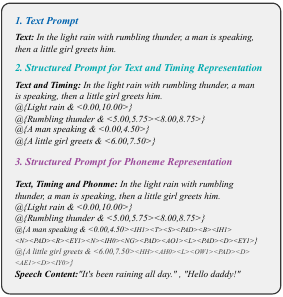}}
\caption{An illustrative example for structured prompt.}
\label{fig:sp}
\end{figure}

\begin{figure*}[htbp]
\centerline{\includegraphics[width=17cm]{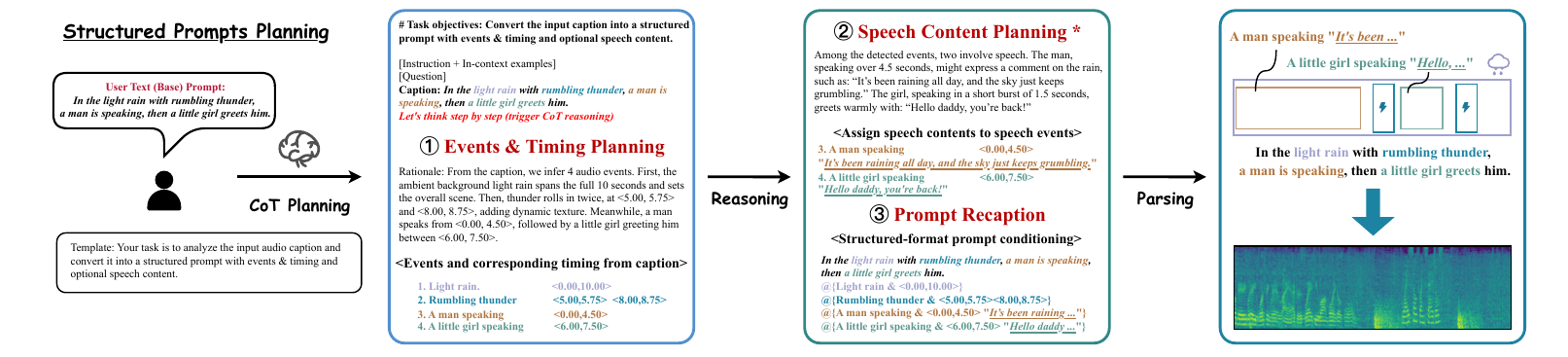}}
\caption{Overview of our CoT-based LLM planning pipeline. Given a user-provided free-form caption, the LLM performs multi-step reasoning to extract audio events with their temporal spans, infer speech content when applicable, and generate a structured prompt that encodes both timing and content for controllable audio generation.}
\label{fig:plan}
\end{figure*}

\subsection{Unified Semantic Modeling}

To address the challenge of encoding diverse condition information, including text, timing, and phoneme features, we propose a unified semantic modeling approach that handles them with a single text encoder in a progressive and coarse-to-fine manner. This approach avoids the complexity of multiple specialized modules~\cite{lee2024voiceldm} by first establishing a robust structural representation, providing a simple yet effective solution for rendering fine-grained content in audio generation.

\noindent\textbf{Structured Prompt for Text and Timing Representation.}
The foundation of our approach is the Structured Prompt ($y_s$), a novel representation we design to explicitly and unambiguously define the composition of an acoustic scene. The prompt employs a standardized format using special tokens to delimit event descriptions and their precise start-and-end times, as illustrated in Figure~\ref{fig:sp}. We propose this format to overcome the critical limitations of using free-form natural language for control. Natural language is often ambiguous; for instance, a prompt like "\textit{an alarm sounds from low to high from 1 second to 9 seconds}" creates confusion, as a model must disentangle whether "\textit{from...to}" refers to a change in pitch or a temporal boundary. Moreover, natural language descriptions become verbose and difficult to parse as scene complexity increases. In contrast, our structured format provides a concise, scalable, and machine-readable representation, forming a robust foundation for generating complex, temporally-aligned audio.

\noindent\textbf{Structured Prompt for Phoneme Representation.}
Our approach to synthesizing intelligible speech is built upon the temporal structure provided by the prompt. A key insight is that the explicit timing windows (<start,end>) assigned to each speech event naturally define the total duration of the utterance. 
This design is consistent with recent TTS studies~\cite{anastassiou2024seed,lee2024ditto,chen2025f5}, which highlight the importance of duration information as a useful constraint for guiding speech generation, either through explicit modeling or by conditioning on the overall utterance duration. Building on this insight, our framework further organizes such temporal information into structured, event-level timing windows, enabling more fine-grained and interpretable control over the generation process.

This simplification makes it natural and highly efficient to use the same, single text encoder to progressively model both the coarse-grained temporal structure and the fine-grained speech content. We therefore represent the speech content at the phoneme level (e.g., "\textit{hello}" $\rightarrow$ [HH, AH0, L, OW1]). Phonemes provide a more direct, pronunciation-aware signal than words, reducing ambiguity and improving the acoustic consistency of the generated speech. By augmenting our single encoder's vocabulary with these phoneme tokens, it learns to render the precise phonetic sequence within the specified temporal boundaries, naturally inheriting the ability to handle speech duration.

\noindent\textbf{Chain-of-Thought (CoT) LLM Planning.}
Given a user-provided free-form description, we employ a CoT-based LLM to convert it into the structured prompt format. The model decomposes the input into temporally-aligned audio events, infers speech content when applicable, and organizes them into a unified representation with explicit timing and phoneme-level information. This process resolves ambiguities in natural language and ensures consistent alignment between semantic content and temporal structure, producing reliable control signals for controllable audio generation.

\subsection{Progressive Model Training}

To train our model for multi-condition audio generation, we adopt a progressive three-stage training strategy. 
This approach allows the model to acquire fine-grained control capability incrementally, where each new stage builds upon and refines the skills learned previously, ensuring a stable and efficient learning process. 
At each stage, the model is optimized using the conditional diffusion objective~\cite{ho2020denoising}, where a network is trained to predict the noise $\epsilon$ added on the clean audio latents:
\begin{equation}
\mathcal{L} = \mathbb{E}_{x, c, \epsilon \sim \mathcal{N}(0, 1), t} \left[ \lVert \epsilon - \epsilon_\theta(z_t, t, \tau_\theta(c)) \rVert_2^2 \right],
\end{equation}
where $z_t$ is the noisy latent at timestep $t$, $\epsilon_\theta$ is the denoising DiT, $c$ is the condition signal, and $\tau_\theta$ is the text encoder. The core of our progressive strategy lies in how the conditioning signal $c$ is structured and utilized across the training stages.

\noindent\textbf{Stage 1: TTA Pre-training.} We first pre-train a DiT~\cite{evans2025stable} on large-scale text-audio datasets to learn a robust, general mapping from textual descriptions to audio latent representation, ensuring~\textit{high-fidelity text-guided audio generation}.

\noindent\textbf{Stage 2: Timing-Controlled TTA Fine-tuning.} The pre-trained model is then fine-tuned on our dataset of precisely timing-annotated audio, while preserving the training on text condition without timing. This stage specifically optimizes the model to interpret the structured prompt containing both text and timing information, achieving~\textit{text-guided and timing-controlled audio generation}.

\noindent\textbf{Stage 3: Timing-Controlled and Intelligible TTA Joint Training.} 
At the final stage, we unfreeze the text encoder to enable joint optimization for both timing control and speech intelligibility. The model is then trained on our full multi-source dataset, which is a comprehensive mixture of our timing-annotated real-world audio and the large-scale simulated data. This final training phase optimizes the model to jointly generate timing-controlled audio and speech samples in a coherent and realistic manner, addressing~\textit{text-guided, timing-controlled, and intelligible audio generation}.

Overall, our progressive model training incrementally acquires finer-grained capabilities while building upon the foundational skills from previous stages. Notably, we find that the joint optimization at Stage 3 not only unlocks speech intelligibility but also further enhances the model's previously learned temporal precision. We attribute these significant improvements to two key factors. The first is the introduction of time-annotated speech data, which provides a richer, more targeted signal for learning the alignment between linguistic content and temporal boundaries. The second is the fine-tuning of the text encoder, which allows it to be jointly optimized with the diffusion backbone; this synergistic training enables both the conditioning (text encoder) and generation (DiT) components to co-adapt to the complex, multi-objective task. This effective co-adaptation is achieved within a simple yet effective framework, where a single text encoder is responsible for processing all conditioning signals: text, timing, and phoneme features.

\begin{table*}
\centering
\caption{Objective and subjective evaluation results on the AudioCondition test set. For each metric, the best result is bold and the second-best is underlined. * denotes models trained by ourselves. $^{\ddagger}$ denotes models evaluated under a different SED model. ControlAudio full denotes evaluation on prompts covering all event classes in the test set.}
\small
\label{tab:timing}
\begin{tabular}{lccccccccc}
\toprule
\multirow[c]{2}{*}[-0.5ex]{Method}  & \multicolumn{2}{c}{\textbf{Temporal (Obj.)}} & \multicolumn{3}{c}{\textbf{Generation (Obj.)}} & \multicolumn{2}{c}{\textbf{Subjective}} & \textbf{Efficiency} \\
\cmidrule(lr){2-3} \cmidrule(lr){4-6} \cmidrule(lr){7-8} \cmidrule(lr){9-9}
 & Eb↑ & At↑ & FAD↓ & KL↓ & CLAP↑ & Temporal↑ & OVL↑ & RTF↓ \\
\midrule
Ground Truth         & 43.37 & 67.53 & -    & -    & 0.377 & 4.52 & 4.48 & -  \\
\midrule
AudioLDM Large           & 6.79  & 35.66 & 3.95 & 2.46 & 0.260 & 1.84 & 2.40 & 1.141 \\
AudioLDM 2 Large         & 7.75  & 42.41 & 3.07 & 1.92 & 0.279 & -    & -    & 1.496 \\
AudioLDM 2 Full Large       & 6.93  & 20.47 & 3.68 & 2.15 & 0.283 & -    & -    & 1.496 \\
Tango                & 1.60  & 26.51 & 2.82 & 1.93 & 0.245 & 1.68 & 2.58 & 1.207 \\
Stable Audio~*       & 11.28 & 51.67 & 1.93 & 1.75 & 0.318 & 1.94 & \underline{3.44} & \underline{0.821} \\
\midrule
CCTA                 & 14.57 & 18.27 & -    & -    & -     & -    & -    & 1.207 \\
MC-Diffusion         & 29.07 & 47.11 & -    & -    & -     & -    & -    & - \\
Tango + LControl     & 21.46 & 55.15 & -    & -    & -     & -    & -    & 1.207 \\
AudioComposer-Small  & 43.51 & 60.83 & 4.92 & 2.00 & 0.261 & 3.12 & 2.52  & \textbf{0.721} \\
AudioComposer-Large  & 44.40 & 63.30 & -    & -    & -     & -    & -     & -     \\
TG-Diff$~^{\ddagger}$& 26.70 & 60.06 & 2.66 & -   & 0.244 & -    & -   & 1.207  \\
FreeAudio            & 44.34 & 68.50 & \underline{1.92} & \underline{1.73} & 0.321 & - & - & 1.166 \\
\midrule
ControlAudio        & \textbf{55.58}  & \textbf{79.52} & 2.61 & 1.85 & \underline{0.325} & \textbf{4.17}    & 3.41 & \underline{0.821} \\
ControlAudio full   & \underline{49.85}  & \underline{71.55} & \textbf{1.47} & \textbf{1.30} & \textbf{0.356} & \underline{3.96}   & \textbf{3.75}  & \underline{0.821} \\
\bottomrule
\end{tabular}
\end{table*}

\subsection{Progressive Guided Sampling}

To optimize our capability to handle both timing and more fine-grained phonetic content, we propose a Progressive Guided Sampling strategy. This approach divides the reverse diffusion process into two phases based on a threshold timestep $t_1$, modulating the conditioning prompt and guidance scale accordingly. Specifically, in the initial sampling phase ($t \in [1.0, t_1]$), we guide the model with a simplified version of our structured prompt that excludes phonetic content $c_1$, using a low guidance scale ($w_{low}$). This encourages the model to first establish a plausible temporal structure for all audio events: 

\vspace{-1mm}
\begin{equation}
    p_{\theta}(z_{t_1:{T-1}}|z_{T},c_1) =\prod_{t=t_1+1}^{T} p_{\theta}(z_{t-1}|z_{t}, c_1).
\end{equation}
\vspace{-1mm}
For the remainder of the sampling process ($t \in (t_1, 0.0]$), we switch to the complete, phoneme-inclusive structured prompt $c_2$ and a higher guidance scale ($w_{high}$). 
\vspace{-1mm}
\begin{equation}
    p_{\theta}(z_{0:{t_1-1}}|z_{t_1},c_2) = \prod_{t=1}^{t_1} p_{\theta}(z_{t-1}|z_t, c_2).
\end{equation}
\vspace{-1mm}

This second phase strictly enforces adherence to the phonetic sequence, ensuring the synthesis of highly intelligible speech within the established structure. This coarse-to-fine strategy improves temporal accuracy and speech clarity by decoupling event placement and content rendering.

\section{Experiments}

\subsection{Experiment Setting} 

\noindent\textbf{Implementation Details.} 
Our model follows a latent diffusion framework with a DiT backbone, where audio is generated in a compressed latent space learned by a DAC-VAE~\cite{evans2024fast}. The autoencoder operates at a 16kHz sampling rate with a latent temporal resolution of 25Hz. 
The diffusion model adopts a DiT architecture following Stable Audio~\cite{evans2024long,evans2025stable}, and generates 10-second audio segments in the latent space. It is conditioned via cross-attention on a pretrained Flan-T5~\cite{chung2024scaling} large text encoder, where text, timing, and phoneme information are unified into a structured prompt and encoded within a shared embedding space. 
We first pretrain a TTA diffusion model for 1M steps, followed by two additional stages of optimization: 0.5M steps for timing-controlled generation and 0.5M steps with the text encoder unfrozen for joint training. Each stage is initialized from the checkpoint of the previous stage, forming a continuous multi-stage training pipeline. All stages are trained on 8 NVIDIA A800 GPUs with a total batch size of 128.
More details on the model architecture and training are provided in the Appendix~\ref{app:model}. Pseudocode for both training and inference procedures is provided in Appendix~\ref{app:pseudocode}.

\noindent\textbf{Evaluation Datasets.} To objectively evaluate our method, we utilize several established datasets, each targeting a specific capability. For timing-controllable generation, we use the publicly available test split from AudioCondition~\cite{guo2024audio}, whose fine-grained temporal annotations are ideal for this task. 
For intelligible speech generation, we use the AC-Filtered~\cite{lee2024voiceldm} dataset. To ensure comparability with prior methods, we rewrite the prompts accordingly, with examples provided in Appendix~\ref{sec:plan-ac}.
For evaluating general TTA performance, we report results on the AudioCaps test set~\cite{kim2019audiocaps}. In addition, we include the LibriTTS-R and LibriSpeech~\cite{panayotov2015librispeech} \textit{test-clean} splits for specific ablation studies.

\noindent\textbf{Evaluation Metrics.} We conduct a comprehensive evaluation covering three key aspects: temporal control, audio quality, and speech intelligibility. For temporal control, we follow prior work~\cite{guo2024audio,wang2025audiocomposer} and report two metrics computed by a sound event detection (SED) system~\cite{mesaros2016metrics}: event-based measures (Eb) and clip-level macro F1 score (At). For audio quality, we employ a suite of standard metrics, including Fréchet Audio Distance (FAD), Kullback–Leibler (KL) divergence, Fréchet Distance (FD), Inception Score (IS)~\cite{liu2023audioldm} and CLAP~\cite{wu2023large}. For speech intelligibility, we conduct both objective and subjective tests. Objectively, we measure the Word Error Rate (WER) by transcribing generated speech with the Whisper \textit{Large-v3} model~\cite{radford2023robust}. Subjectively, we conduct Mean Opinion Score (MOS) tests where 20 participants rate three aspects on a five-point scale: Speech Intelligibility, Overall Quality (OVL), and Relevance to the prompt (REL). Further details are provided in the Appendix~\ref{sec:eval}.

\begin{table*}
\centering
\small
\caption{Objective and subjective evaluation results on the AC-Filtered.}
\label{tab:tts}
\begin{tabular}{lccccccccc}
\toprule
\multirow[c]{2}{*}[-0.5ex]{Method}  & \multicolumn{6}{c}{\textbf{Objective}} & \multicolumn{3}{c}{\textbf{Subjective}} \\
\cmidrule(lr){2-7} \cmidrule(lr){8-10}
 & FAD↓ & KL↓ & FD↓ & IS↑ & CLAP↑ & WER↓ & Intelligible↑ & OVL↑ & REL↑ \\
\midrule
Ground Truth      & -     & -    & -      & -     & 0.523  & 17.47                                                    & 4.16 & 4.45 & 4.50   \\
\midrule
AudioLDM 2 Speech & 23.55 & 3.58 & 102.84 & 1.52  & 0.078 & 32.74                                                     & 2.85 & 1.92 & 1.60    \\
VoiceLDM-S        & 4.46  & 1.52 & 47.08  & \underline{3.40}  & \underline{0.479} & 43.21                             & 2.62 & 2.55 & 2.51    \\
VoiceLDM-M        & 5.90  & \textbf{1.43} & \underline{46.40}  & 3.16  & 0.458 & 8.84                              & \underline{4.18} & \underline{3.64} & \underline{3.47}    \\
VoiceDiT          & \underline{4.60}  & -    & -      & -     & 0.220 & \underline{7.09}   & -       & -       & -    \\
\midrule
ControlAudio      & \textbf{3.52} & \underline{1.45} & \textbf{32.55} & \textbf{4.43}  & \textbf{0.513} & \textbf{6.84}  & \textbf{4.31} & \textbf{4.15} & \textbf{3.82}    \\
\bottomrule
\end{tabular}
\end{table*}

\subsection{Main Results}

\noindent\textbf{Timing-Controlled Audio Generation.} 
We compare ControlAudio with several state-of-the-art TTA models, including AudioLDM~\cite{liu2023audioldm}, AudioLDM 2~\cite{liu2024audioldm}, Tango~\cite{ghosal2023text}, and our in-house version of Stable Audio~\cite{evans2024fast,evans2024long,evans2025stable}. We further include methods with explicit temporal conditioning, such as MC-Diffusion~\cite{guo2024audio} and AudioComposer~\cite{wang2025audiocomposer}, as well as training-free baselines TG-Diff~\cite{du2024controllable} and FreeAudio~\cite{jiang2025freeaudio}, covering a broad range of paradigms for controllable TTA generation. 
TG-Diff reports both timing and audio quality metrics under a training-free framework, but relies on a different sound event detection model~\cite{turpault2019sound} from other baselines. CCTA is a variant of MC-Diffusion that uses only control conditions without textual input, while Tango + LControl extends Tango with language-based temporal control. For AudioComposer, we use its publicly available Small variant for evaluation.
We evaluate all methods from three perspectives: temporal alignment, audio quality, and efficiency. For efficiency, we measure the real-time factor (RTF)~\cite{liu2024audiolcm,liu2024flashaudio} on a single NVIDIA A800 GPU. As shown in Table~\ref{tab:timing}, ControlAudio achieves competitive or superior temporal alignment compared to existing methods, while consistently improving audio quality in both objective and subjective metrics. A detailed category-wise breakdown on AudioCondition is provided in Appendix~\ref{app:audiocondition_details}, reporting Eb and At metrics across different event categories. Importantly, these improvements are achieved without introducing additional inference overhead, demonstrating a favorable trade-off between controllability, quality, and efficiency.

\noindent\textbf{Intelligible Audio Generation.} We further assess the ability of ControlAudio to generate intelligible speech on the AC-Filtered, comparing it with speech-oriented baselines including AudioLDM 2 Speech, VoiceLDM-S, VoiceLDM-M, and VoiceDiT. To evaluate these baselines lacking native timing support, we first use an LLM to predict a plausible time window from the caption, with further details in the Appendix~\ref{sec:plan-ac}. For this comparison, we use the publicly available checkpoints of VoiceLDM, while directly reporting the results presented in the original VoiceDiT paper.
As shown in Table~\ref{tab:tts}, ControlAudio achieves lower WER, superior audio quality, and improved text-audio semantic alignment compared to all baselines.
Subjective evaluations also indicate improvements in speech intelligibility, overall audio quality, and text relevance, demonstrating that ControlAudio can generate clearer and more faithful speech segments while preserving general audio fidelity.

\noindent\textbf{Text-to-Audio Generation.} To verify that introducing timing and speech content control does not compromise general TTA generation capabilities, we evaluate ControlAudio on the AudioCaps test set under standard natural language captions, comparing against both text-to-audio models and other controllable audio generation baselines. Unlike prior controllable generation approaches that often sacrifice audio quality for control precision, ControlAudio maintains high generative performance while providing fine-grained controllability. As shown in Table~\ref{tab:tta}, ControlAudio achieves competitive or superior results across multiple audio quality metrics compared to state-of-the-art baselines. These findings demonstrate that our structured prompt conditioning and vocabulary extension can be seamlessly integrated into a T2A system, enabling precise timing and intelligible speech control without degrading semantic alignment or acoustic fidelity.

\begin{table}
\centering
\scriptsize
\caption{Objective and subjective evaluation results on the AudioCaps test set.}
\label{tab:tta}
\begin{tabular}{lccccc}
\toprule
Method & FAD↓ & KL↓ & FD↓ & IS↑ & CLAP↑ \\
\midrule
Ground Truth          & -                & -    & -     & -     & 0.525 \\
\midrule
AudioGen              & 1.82             & 1.69 & -     & -     & -     \\
AudioLDM              & 4.96             & 2.17 & 29.29 & 8.13  & 0.373 \\
AudioLDM 2            & 2.12             & 1.54 & 33.18 & 8.29  & 0.281 \\
Tango                 & 1.73             & \underline{1.27} & 24.42 & 7.70  & 0.315 \\
Tango 2               & 2.63             & \textbf{1.12} & 20.66 & 9.09  & 0.375 \\
Stable Audio~*        & \textbf{1.52}    & 1.51 & \underline{18.30} & \underline{13.79} & \textbf{0.538} \\
\midrule
AudioComposer-S   & 3.63             & 1.76 & 27.57 & -     & -     \\
AudioComposer-L   & 2.52             & 1.39 & 19.25 & -     & -     \\
VoiceLDM-S            & 13.83            & 3.36 & 63.42 & 4.56  & 0.217 \\
VoiceLDM-M            & 9.70             & 2.81 & 55.80 & 4.60  & 0.272 \\
VoiceDiT              & 3.55             & 1.87 &  -    & -     & 0.450 \\
\midrule
ControlAudio          & \underline{1.56} & 1.31 & \textbf{14.20} & \textbf{14.49} & \underline{0.535} \\
\bottomrule
\end{tabular}
\end{table}

\subsection{Ablation Study}

\noindent\textbf{Ablation of Prompt Design.} To isolate and evaluate the effectiveness of our structured prompt design, we conduct a targeted ablation study. For this analysis, we compare a baseline model trained with conventional natural language descriptions against our model trained with structured prompts. Crucially, both models are trained only up to Stage 2 of our progressive curriculum, the phase dedicated specifically to learning timing control. This controlled setting allows us to fairly assess the impact of the prompt format itself. As shown in Table~\ref{tab:prompt}, the model trained with structured prompts consistently achieves superior temporal alignment and overall audio quality on the AudioCondition test set. The results suggest that the structured format provides a clearer, unambiguous mapping between events and their time spans, an advantage that becomes particularly pronounced in complex scenes where verbose natural language descriptions can degrade timing accuracy.

\begin{table}[t]
\centering
\footnotesize
\caption{Comparison of prompt formats on the AudioCondition test set. We compare Natural Language (NL) with our Structured Prompt (SP).}
\label{tab:prompt}
\begin{tabular}{lccccc}
\toprule
Format   & Eb↑   & At↑   & FAD↓ & KL↓           & CLAP↑   \\
\midrule
NL       & 46.23 & 65.36 & 4.11 & 2.25 & 0.245\\
SP       & \textbf{51.62} & \textbf{70.81} & 3.61 & 2.05 & 0.293\\
NL full  & 40.79 & 61.06 & 1.03 & 1.36 & 0.376 \\
SP full  & 43.76 & 64.82 & \textbf{0.92} & \textbf{1.27} & \textbf{0.419}\\
\bottomrule
\end{tabular}
\end{table}

\noindent\textbf{Ablation of Vocabulary Granularity.} To determine the optimal vocabulary granularity for intelligible speech, we conduct an ablation study on the LibriTTS-R and LibriSpeech test-clean datasets. We first compare three variants of our model, differentiated by their vocabulary: word-level, sub-word (BPE), and phoneme-level. For broader context, we also report results from the strong VoiceLDM baselines. Evaluation metrics include WER for intelligibility and UTMOS~\cite{saeki2022utmos}(UT-M) for speech naturalness. For WER calculation on LibriSpeech, we adopt a HuBERT-based ASR model~\cite{hsu2021hubert}, following prior works~\cite{shen2023naturalspeech}. As shown in Table~\ref{tab:token}, the phoneme-level model consistently and significantly outperforms the other granularities, achieving the lowest WER and highest UTMOS scores. These findings confirm that phonemes provide a more direct representation of spoken content, facilitating a tighter alignment between the prompt and the acoustic output, which ultimately translates into substantially clearer and more intelligible speech.

\begin{table}[t]
\small
\footnotesize
\caption{Comparison of vocabulary extension strategies for intelligible speech synthesis on LibriTTS-R and LibriSpeech test sets.}
\label{tab:token}
\begin{tabular}{lcccc}
\toprule
\multirow[c]{2}{*}[-0.5ex]{Token Type} & \multicolumn{2}{c}{LibriTTS-R} & \multicolumn{2}{c}{LibriSpeech} \\
\cmidrule(lr){2-3} \cmidrule(lr){4-5}
                      & WER↓  & UT-M↑  & WER↓  & UT-M↑ \\
\midrule
Ground Truth          & 3.75  & 4.17    & 2.15  & 4.06 \\
\midrule
VoiceLDM-S            & 36.65 & 2.59    & 38.61 & 2.76 \\
VoiceLDM-M            & 4.98  & 2.83    & 9.76  & 2.77 \\
\midrule
Word                  & 6.96  & 4.12    & 6.44  & 4.14 \\
BPE                   & 7.53  & 4.15    & 5.04  & 4.20 \\
Phoneme               & \textbf{4.00}  & \textbf{4.18}    & \textbf{3.62}  & \textbf{4.22} \\
\bottomrule
\end{tabular}
\end{table}

\begin{figure}[t]
\centering
\centerline{\includegraphics[width=7.5cm]{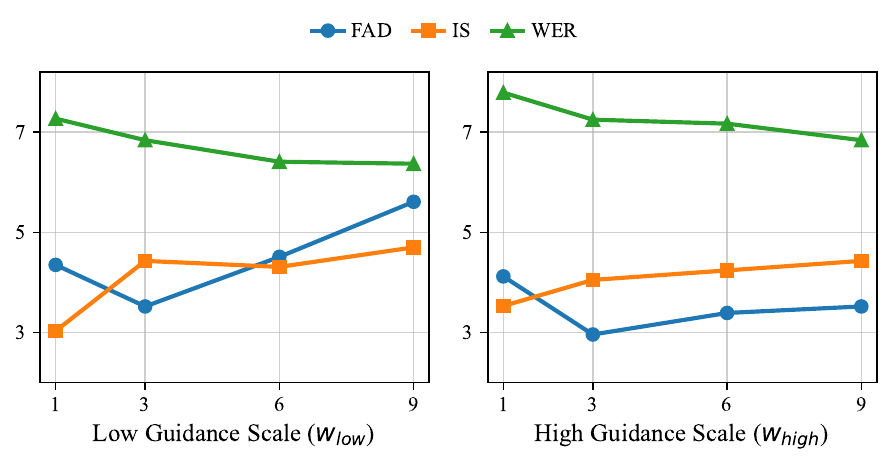}}
\caption{Analysis of Progressive Sampling parameters ($w_{low}$, $w_{high}$). This study reveals a clear trade-off between audio quality and speech intelligibility.}
\label{fig:cfg}
\end{figure}

\noindent\textbf{Analysis of Sampling Strategy.}
We conduct an analysis to validate our progressive sampling strategy. This coarse-to-fine approach first uses a low guidance scale ($w_{low}$) with a simplified, content-free prompt to establish the temporal structure. It then transitions to a high scale ($w_{high}$) with the full, phoneme-inclusive prompt to render intelligible speech.
As visualized in Figure~\ref{fig:cfg}, our analysis of varying $w_{low}$ and $w_{high}$ reveals a clear trade-off: a low initial scale is crucial for overall audio quality, while a high subsequent scale is essential for speech intelligibility. This study empirically identifies the optimal configuration as ($w_{low}=3, w_{high}=9$), confirming the effectiveness of our approach.
For a total of $T=100$ sampling steps, we set the transition timestep to $t_1 = 88$. We also examine the sensitivity of $t_1$: a smaller $t_1$ (longer low-guidance phase) improves audio quality, while a larger $t_1$ reduces the steps for phoneme refinement and harms intelligibility. The performance remains stable across a reasonable range of $t_1$. Detailed sensitivity results of $t_1$ are provided in Appendix~\ref{app:cfg_t1}.

\section{Conclusion}

In this work, we introduced ControlAudio, which recasts controllable TTA generation as a multi-task learning problem solved via a progressive diffusion modeling strategy. This progressive approach is applied across data construction, model training, and inference, enabling our model to incrementally master fine-grained control from text, timing, and phoneme conditions. Extensive experiments demonstrate that ControlAudio achieves state-of-the-art performance in both temporal accuracy and speech clarity. Our work's potential for misuse in creating deceptive content or voice impersonations underscores the urgent need for robust detection methods and responsible AI governance.

\section{Acknowledgments}
This work is supported by the Fundamental and Interdisciplinary Disciplines Breakthrough Plan of the Ministry of Education of China (No. JYB2025XDXM101), and the National Natural Science Foundation of China (62550004, U24A20342, U25B6003, 92570001).

\section*{Limitations}

Despite its promising results, our work has several limitations. First, while ControlAudio pioneers the generation of intelligible speech within a timing-controlled TTA framework, its control is primarily limited to the speech content. The framework currently lacks explicit mechanisms to manipulate crucial stylistic attributes such as emotion, prosody, or speaker identity. Second, a fundamental tension between generating high-quality general audio versus intelligible speech persists. Although our model unifies these tasks, we observe a potential trade-off where heavily optimizing for one modality can slightly impact the fidelity of the other in complex, co-occurring scenes. Finally, the performance of our model is inherently constrained by the availability of large-scale, richly annotated audio-speech datasets, which remain scarce. Our reliance on a combination of existing annotated data and simulated data, while effective, suggests that performance could be further enhanced with the advent of more comprehensive and higher-quality training corpora in the future.

\bibliography{acl_latex}

\clearpage

\appendix

\section{Training Datasets}
\label{sec:train_data}

\subsection{Pretraining Datasets and Preprocessing}
\label{sec:pre_data}

Table~\ref{tab:dataset} provides a comprehensive summary of all corpora used for pretraining our TTA backbone. To learn a robust mapping between text and audio, we aggregate a diverse mixture of large-scale, publicly available datasets. This includes datasets with descriptive captions, such as the large-scale WavCaps~\cite{mei2024wavcaps} and the widely-used AudioCaps~\cite{kim2019audiocaps}, as well as corpora with high-level event labels, like the massive AudioSet~\cite{gemmeke2017audio}. This rich combination of data sources, spanning both detailed descriptions and a wide vocabulary of sound classes, allows the model to learn robust and versatile semantic representations.

All audio samples from these sources undergo a standardized preprocessing pipeline. First, all audio is resampled to 16kHz and converted to a mono-channel format. To accommodate the fixed-size input requirement of our diffusion model, all clips are processed into a uniform 10-second duration. Samples shorter than 10 seconds are right-padded with silence, while for samples longer than 10 seconds, a random 10-second segment is cropped.

\begin{table}[htbp]
\centering
\caption{Details about audio-text datasets we use.}
\label{tab:dataset}
\begin{tabular}{cccc}
\toprule
\textbf{Dataset} & \textbf{Hours}(h)  & \textbf{Number}  & \textbf{Text}  \\
\midrule
AudioCaps  & 109   & 44K  & caption \\
WavCaps    & 7090  & 400K & caption \\
Clotho v2  & 152   & 7k   & caption \\
AudioSet   & 5800  & 2M   & label   \\
FSD50k     & 108   & 51K  & label   \\
ESC-50     & 2.8   & 2K   & label   \\
VGG-Sound  & 550   & 210k & label   \\
\hline
MTT        & 200   & 24K  & caption\\
MSD        & 7333  & 880K & caption\\
FMA        & 900   & 11K  & caption\\
\bottomrule
\end{tabular}
\end{table}

\subsection{Timing-Controlled Datasets}
\label{sec:timing_data}

For the timing-control fine-tuning stage, our dataset is constructed based on AudioSet-Strong~\cite{hershey2021benefit}, which contains 1.8M audio clips. This dataset is crucial as it provides dense, frame-level timestamps for 456 sound event classes. However, since AudioSet-Strong only provides categorical labels (e.g., "Dog"), not descriptive text, we generate richer captions for each timed event. Inspired by the methodology of WavCaps~\cite{mei2024wavcaps}, we employ a large language model (LLM) to create a unique textual description for each segmented audio event.

A critical difference in preprocessing this dataset is the handling of audio duration to preserve the integrity of the timestamps. Unlike in the pretraining phase, we do not apply random cropping. Instead, we consistently take the first 10 seconds of each audio clip and subsequently filter the event annotations, retaining only those whose timestamps fall within this 0-10s window. Clips shorter than 10 seconds are right-padded with silence. This deterministic process ensures that the temporal annotations in our final dataset remain perfectly aligned with the corresponding audio segments.

\subsection{Annotated Data Pipeline}
\label{sec:anno_data}

This section provides a detailed, step-by-step description of the pipeline used to create our annotated dataset of real-world speech events with both precise temporal boundaries and textual transcriptions. The process is as follows:

\noindent \textbf{Initial Data Selection from AudioSet-SL.} We begin with the AudioSet-SL dataset~\cite{hershey2021benefit}, which contains strong, human-verified temporal annotations for a wide range of sound events. From the full dataset, we first identify and select all 10-second audio clips that contain at least one event labeled as "Human speech," "Speech," or any of their subcategories. This initial filtering yields a subset of 49,950 clips containing speech.

\noindent \textbf{High-Quality Speech Track Extraction.} To obtain high-quality and reliable speech stems, we employ a dual-demixing comparison strategy inspired by MTV~\cite{weng2025audio}. This strategy involves comparing the separation outputs from MVSEP~\cite{ZFTurbo23} and Spleeter~\cite{hennequin2020spleeter} to filter for quality and extract a clean speech signal for the subsequent processing steps.

\noindent \textbf{Event-Level Segmentation.} The extracted clean speech track is then segmented into individual, non-overlapping speech events. We use the original, human-annotated start and end timestamps provided by AudioSet-SL to perform this segmentation. Each resulting audio segment represents a single, continuous speech utterance from the original recording. This process yields a total of 173,831 individual speech segments, which are then prepared for transcription.

\noindent \textbf{Transcription with Large Language Model.} Each of the 173,831 clean, segmented speech events is then sent for transcription. We input the audio segment into the Gemini 2.5 Pro model with a direct prompt to generate a precise textual transcription. To ensure the quality of the final annotations, we explicitly instruct the model to return an empty output if the spoken content in an audio segment is unintelligible or heavily obscured by noise. This step serves as a crucial quality filter.

This entire pipeline, from segmentation to filtered transcription, results in our final annotated dataset. From the initial pool of segments, a total of 152,070 high-quality, transcribed speech events are retained. Each event in this dataset is characterized by a precise start time, end time, and a verified textual transcription, providing an authentic and challenging data source for training our model on real-world, timed speech.

\subsection{Simulated Data Pipeline}
\label{app:simu_data}

In addition to the annotated real-world data, we developed a pipeline to construct a large-scale simulated dataset. The goal of this pipeline is to generate realistic, complex audio scenes with precise timing and transcription information, guided by the statistical patterns observed in a real-world dataset. The process consists of two main stages: deriving statistical priors and the guided synthesis itself.

\noindent\textbf{Deriving Statistical Priors from AudioSet-SL.} To ensure our simulated data reflects real-world patterns of speech activity, we first perform a statistical analysis on the speech-containing clips within AudioSet-SL. We identify two key distributions:
\begin{itemize}[leftmargin=*]
\item \textbf{Speaker Distribution:} We find that approximately 79.1\% of clips (39,509 out of 49,950) feature a single speaker, while 20.9\% feature multiple speakers. This ratio guides the proportion of monologue vs. dialogue scenarios in our simulation.
\item \textbf{Utterance-per-Clip Distribution:} The empirical distribution is characterized by a prominent peak at a single utterance per clip ($n=1$), which accounts for 32.20\% of all single-speaker scenarios. For $n > 1$, the frequency of clips generally decreases as the number of utterances increases, exhibiting a long tail. We sample from this distribution to determine the number of utterances in our simulated monologues, with the maximum number of utterances per clip capped at 8 to focus on the most prevalent scenarios. The full distribution is provided in Table~\ref{tab:utterance_dist}.
\end{itemize}

\begin{table}[htbp]
\centering
\caption{Empirical distribution of the number of utterances per 10-second single-speaker clip, analyzed from AudioSet-SL. This distribution guides the synthesis of our simulated monologue data.}
\label{tab:utterance_dist}
\begin{tabular}{ccc}
\toprule
\textbf{Events ($n$)} & \textbf{Number} & \textbf{Percentage (\%)} \\
\midrule
1  & 12,723 & 32.20 \\
2  & 6,462  & 16.36 \\
3  & 6,284  & 15.90 \\
4  & 5,720  & 14.48 \\
5  & 4,201  & 10.63 \\
6  & 2,328  & 5.89  \\
7  & 1,047  & 2.65  \\
8  & 456    & 1.15  \\
\midrule
9  & 150    & 0.38  \\
10 & 67     & 0.17  \\
11 & 41     & 0.10  \\
12 & 20     & 0.05  \\
$n > 12$ & 10      & 0.04 \\
\midrule
\textbf{Total} & \textbf{39,509} & \textbf{100.00} \\
\bottomrule
\end{tabular}
\end{table}

\noindent\textbf{Guided Synthesis Pipeline.} The synthesis process for each 10-second clip is as follows. First, we determine the scenario type by sampling from the speaker distribution (a 79.1\% chance of a single-speaker monologue). Next, we source clean speech utterances with transcripts from the LibriTTS-R dataset~\cite{koizumi2023libritts}. For a monologue, we sample a number of utterances (determined by the utterance-per-clip distribution, and capped at a maximum of 8) from a single speaker. For a dialogue, we sample utterances from 2 to 4 different speakers, ensuring that no single speaker contributes more than 4 utterances. We then simulate a plausible temporal arrangement for these utterances within the 10-second window. Finally, the composed speech-only track is mixed with a non-speech background audio clip randomly selected from a filtered subset of WavCaps~\cite{mei2024wavcaps} and VGG-Sound~\cite{chen2020vggsound}. The mixing is performed at a signal-to-noise ratio (SNR) randomly sampled from a uniform distribution between 2 and 10~dB.

\section{Model Configurations}
\label{app:model}
This section details the architecture of the base model used for pretraining, before it is fine-tuned into ControlAudio. Our diffusion model is built upon the DiT (Diffusion Transformer) architecture within a latent diffusion modeling (LDM) paradigm. For pretraining, the model is conditioned on three input types: a natural language prompt (\texttt{prompt}), the start time (\texttt{seconds\_start}), and the total duration (\texttt{seconds\_total}). All conditions are embedded into a 768-dimensional feature space. The prompt is encoded using a pretrained Flan-T5 large model, while \texttt{seconds\_start} and \texttt{seconds\_total} are treated as numerical inputs.

The diffusion network backbone is a DiT with 24 layers, 24 attention heads, and a model hidden dimension of 1536~\cite{evans2024long}. The model utilizes both cross-attention for all conditional inputs and global conditioning for duration-related signals. The internal token dimension of the diffusion model is 64, with a conditional token dimension of 768 and a global condition embedding dimension of 1536.

\subsection{Compression Networks}

Our audio autoencoder is a variational autoencoder (VAE) based on the Descript Audio VAE~\cite{evans2025stable} framework, operating at a 16kHz sampling rate. The model is trained from scratch on the audio portions of large-scale public datasets to learn a compact audio representation. The encoder is configured with a model dimension (\texttt{d\_model}) of 128 and uses strides of [4, 4, 4, 10], resulting in an overall downsampling ratio of 640. The encoder maps the input waveform into a final 64-dimensional latent representation, which is then used by the decoder for reconstruction. The model's input/output channels (\texttt{io\_channels}) are set to 1 for mono audio. We use Snake activation throughout the network and omit the final tanh activation in the decoder.

\subsection{Training Details}

To improve convergence stability and generation quality, we adopt several common training strategies~\cite{evans2025stable}, with configurations specified in our training setup. We apply Exponential Moving Average (EMA) to the model parameters. For optimization, we use the AdamW optimizer with a learning rate of $5 \times 10^{-5}$, $(\beta_1, \beta_2) = (0.9, 0.999)$, and a weight decay of $1 \times 10^{-3}$. 

Our learning rate schedule consists of two phases. For the first 99\% of training iterations, the learning rate is held constant at its initial value, $\eta_0$. For the final 1\% of iterations, it then decays following the InverseLR formula:
\begin{equation}
\eta_t = \eta_0 \times \left(1 + \frac{t'}{\gamma}\right)^{-\text{power}},
\end{equation}
where $t'$ is the step count within the decay phase, $\gamma = 10^6$, and $\text{power} = 0.5$. This strategy allows for stable and rapid convergence in the main training phase, followed by a short period of fine-tuning with a decaying learning rate.

In this final stage, we initialize the model from the Stage 2 checkpoint and unfreeze the Flan-T5 text encoder, enabling joint optimization with the diffusion backbone. The optimization configurations are retained from the previous stages. This joint training is crucial as it allows the text encoder to adapt its representations to the composite nature of our prompt, which includes the structured format, special tokens for timing, and the extended phoneme-level vocabulary for speech. As a result, the model learns a unified representation that maps diverse inputs, such as semantic descriptions, precise temporal spans, and intelligible speech content, to a single, high-quality, timing-controlled audio output.

\section{Evaluation}
\label{sec:eval}

\subsection{Objective Metrics}

We conduct a comprehensive objective evaluation to assess our model's performance in two key areas: audio quality and semantic alignment with the text prompt.

\noindent\textbf{Audio Quality.} Our primary metric for audio fidelity is the Fréchet Audio Distance (FAD), which measures the distributional difference between generated and reference audio based on VGGish embeddings. To evaluate the consistency of acoustic event distributions, we also report Kullback-Leibler (KL) divergence computed using the PANNs tagging model. For completeness and comparison with prior works, we include the Inception Score (IS) and Fréchet Distance (FD) as supplementary metrics~\cite{liu2023audioldm}.

\noindent\textbf{Semantic Alignment.} To measure the alignment between the generated audio and its corresponding text prompt, we use the LAION-CLAP~\cite{wu2023large} score. This score is defined as the cosine similarity between the CLAP embeddings of the generated audio $a$ and the text prompt $t$:
\begin{equation}
\text{CLAP}(a, t) = \frac{a \cdot t}{\|a\| \, \|t\|}.
\end{equation}
A higher CLAP score indicates better semantic correspondence in the shared embedding space. All objective metrics are computed using the official AudioLDM evaluation toolkit for consistency.

\subsection{Subjective Evaluation}

For our subjective evaluation, we recruited 20 human evaluators to rate generated audio samples on a 5-point Mean Opinion Score (MOS) scale (1-5, with higher scores being better). The evaluation was divided into two distinct tasks, each with specific criteria:

\noindent\textbf{Timing-Controlled Audio Generation.} In this task, participants were presented with an audio clip and its corresponding timed prompt. They were asked to rate the audio based on the following two aspects:
\begin{itemize}[leftmargin=*]
    \item \textbf{Temporal Alignment (Temporal):} This measures the accuracy of timestamp adherence. The question asked was: "How accurately does the timing of the audio events match the given start and end times in the prompt?"
    \item \textbf{Overall Quality (OVL):} This assesses the perceptual quality of the audio clip itself. The question asked was: "Ignoring the prompt, how would you rate the overall quality and realism of the audio clip?"
\end{itemize}

\noindent\textbf{Intelligible Audio Generation.} In this task, participants were presented with an audio clip containing speech and the text it was intended to convey. They rated the audio based on the following aspects:
\begin{itemize}[leftmargin=*]
    \item \textbf{Speech Intelligibility (Intelligible):} This measures the clarity of the spoken content. The question asked was: "How clear and understandable is the spoken content in the audio?"
    \item \textbf{Overall Quality (OVL):} This assesses the quality of the entire acoustic scene. The question asked was: "How would you rate the overall audio quality, including both the speech and any background sounds?"
    \item \textbf{Relevance (REL):} This measures the semantic correspondence between the audio and the text. The question asked was: "How well does the generated audio, as a whole, match the text description?"
\end{itemize}

\section{Training and Inference Pseudocode}
\label{app:pseudocode}

Algorithms~\ref{alg:training} and~\ref{alg:inference} summarize the training and inference procedures of ControlAudio. The training process follows our progressive three-stage design, where the model gradually expands its controllability from text, to timing, and finally to phoneme-level speech content. During training, different condition combinations are randomly switched to preserve previously learned capabilities while introducing finer-grained control signals. During inference, we adopt a two-phase progressive sampling strategy, where temporal structure is first established with low guidance and the phoneme condition is then introduced with higher guidance to refine intelligible speech.

\begin{algorithm}[h]
\caption{Progressive Training}
\label{alg:training}
\small
\KwIn{Training datasets $\mathcal{D}_1, \mathcal{D}_2, \mathcal{D}_3$}
\KwOut{Trained ControlAudio model $\epsilon_\theta$}

Initialize diffusion model $\epsilon_\theta$ and a pretrained text encoder $\tau_\theta$\;

\textbf{Stage 1: Text-to-Audio Pretraining}\;
\For{each training step}{
    Sample $(x, c_{\text{text}}) \sim \mathcal{D}_1$\;
    Sample $t \sim \mathcal{U}(0,1), \epsilon \sim \mathcal{N}(0,1)$\;
    $z_t \leftarrow \text{ForwardDiffusion}(x, t, \epsilon)$\;
    $\hat{\epsilon} \leftarrow \epsilon_\theta(z_t, t, \tau_\theta(c_{\text{text}}))$\;
    Update $\theta$ with $\|\epsilon - \hat{\epsilon}\|^2$\;
}

\textbf{Stage 2: Timing-Controlled Training}\;
\For{each training step}{
    Sample $(x, c_{\text{text}}, c_{\text{timing}}) \sim \mathcal{D}_2$\;
    Randomly choose condition $c \in \{c_{\text{text}}, (c_{\text{text}}, c_{\text{timing}})\}$\;
    Sample $t, \epsilon$\;
    $z_t \leftarrow \text{ForwardDiffusion}(x, t, \epsilon)$\;
    $\hat{\epsilon} \leftarrow \epsilon_\theta(z_t, t, \tau_\theta(c))$\;
    Update $\theta$\;
}

\textbf{Stage 3: Joint Training with Phoneme}\;
Unfreeze text encoder $\tau_\theta$\;
\For{each training step}{
    Sample $(x, c_{\text{text}}, c_{\text{timing}}, c_{\text{phoneme}}) \sim \mathcal{D}_3$\;
    Randomly choose condition $c \in \{c_{\text{text}}, (c_{\text{text}}, c_{\text{timing}}), (c_{\text{text}}, c_{\text{timing}}, c_{\text{phoneme}})\}$\;
    Sample $t, \epsilon$\;
    $z_t \leftarrow \text{ForwardDiffusion}(x, t, \epsilon)$\;
    $\hat{\epsilon} \leftarrow \epsilon_\theta(z_t, t, \tau_\theta(c))$\;
    Update $\theta$\;
}
\end{algorithm}

\begin{algorithm}[h]
\caption{Progressive Sampling}
\label{alg:inference}
\small
\KwIn{Text prompt $c$, timestep threshold $t_1$, guidance scales $w_{\text{low}}, w_{\text{high}}$}
\KwOut{Generated audio sample $x$}

Initialize $z_T \sim \mathcal{N}(0, I)$\;

Construct:
$c_1 = (c_{\text{text}}, c_{\text{timing}})$\;
$c_2 = (c_{\text{text}}, c_{\text{timing}}, c_{\text{phoneme}})$\;

\For{$t = T$ to $t_1$}{
    $\hat{\epsilon} \leftarrow \epsilon_\theta(z_t, t, c_1)$\;
    Apply classifier-free guidance with scale $w_{\text{low}}$\;
    $z_{t-1} \leftarrow \text{Denoise}(z_t, \hat{\epsilon})$\;
}

\For{$t = t_1$ to $0$}{
    $\hat{\epsilon} \leftarrow \epsilon_\theta(z_t, t, c_2)$\;
    Apply classifier-free guidance with scale $w_{\text{high}}$\;
    $z_{t-1} \leftarrow \text{Denoise}(z_t, \hat{\epsilon})$\;
}

Return generated audio $x = \mathrm{Decode}(z_0)$\;
\end{algorithm}

\begin{table*}[t]
\centering
\small
\setlength{\tabcolsep}{4pt}
\caption{Category-wise performance on the AudioCondition test set.}
\label{tab:audiocondition_details}

\begin{tabular}{lccccccccccc}
\toprule
\multicolumn{12}{l}{\textbf{Eb}$\uparrow$} \\
\midrule
Method & Alarm & Blender & Cat & Dishes & Dog & T-brush & Frying & Water & Speech & Cleaner & Avg \\
\midrule
FreeAudio & 40 & 47 & 14 & 20 & 23 & 78 & 60 & 51 & 39 & 72 & 44.34 \\
ControlAudio & 53.08 & 48.57 & 27.06 & 40.55 & 50.41 & 80.51 & 61.15 & 52.01 & 64.23 & 78.26 & 55.58 \\
\midrule
\multicolumn{12}{l}{\textbf{At}$\uparrow$} \\
\midrule
Method & Alarm & Blender & Cat & Dishes & Dog & T-brush & Frying & Water & Speech & Cleaner & Avg \\
\midrule
FreeAudio & 73 & 67 & 38 & 42 & 59 & 88 & 85 & 81 & 76 & 76 & 68.50 \\
ControlAudio & 87.22 & 69.39 & 41.25 & 67.48 & 85.50 & 91.67 & 91.38 & 82.69 & 88.65 & 90.00 & 79.52 \\
\bottomrule
\end{tabular}
\end{table*}

\begin{table}[t]
\centering
\small
\caption{Sensitivity analysis of the transition timestep $t_1$ (coarse range).}
\label{tab:cfg_t1_coarse}
\begin{tabular}{l|ccccc}
\toprule
$t_1$ & 100 & 95 & 90 & 85 & 80 \\
\midrule
FAD$\downarrow$ & 5.82 & 4.74 & 3.85 & 3.40 & 3.35 \\
WER$\downarrow$ & 5.98 & 6.61 & 6.83 & 7.12 & 7.67 \\
\bottomrule
\end{tabular}
\end{table}

\begin{table}[t]
\centering
\small
\caption{Sensitivity analysis of the transition timestep $t_1$ (fine-grained range).}
\label{tab:cfg_t1_fine}
\begin{tabular}{c|cccccc}
\toprule
$t_1$ & 90 & 89 & 88 & 87 & 86 & 85 \\
\midrule
FAD$\downarrow$ & 3.85 & 3.61 & 3.52 & 3.53 & 3.47 & 3.40 \\
WER$\downarrow$ & 6.83 & 6.85 & 6.84 & 6.86 & 7.01 & 7.12 \\
\bottomrule
\end{tabular}
\end{table}

\section{Detailed Results}

\subsection{Category-wise Results on AudioCondition}
\label{app:audiocondition_details}

For the AudioCondition dataset, we provide detailed category-wise results in Table~\ref{tab:audiocondition_details}, including both event-based (Eb) and clip-level (At) metrics. ControlAudio consistently improves performance across most categories, particularly on speech-related and temporally complex events.

\subsection{Sensitivity Analysis of Transition Timestep $t_1$}
\label{app:cfg_t1}

A clear trade-off between audio quality and speech intelligibility is observed: smaller $t_1$ improves FAD, while larger $t_1$ benefits WER. A fine-grained analysis around the selected setting further shows that the performance remains stable within a moderate range. Based on this trade-off, we set $t_1 = 88$ in all experiments.

\section{LLM Planning}

\subsection{Chain-of-Thought for Prompt Planning}

Recent advances have demonstrated the powerful planning and cross-modal reasoning capabilities of large language models (LLMs)~\cite{wang2025multimodal,wang2024qwen2}. We leverage these capabilities by employing an LLM to function as a "planner" that automatically converts a free-form natural language caption ($y_c$) into a precise, structured prompt ($y_s$) for our generative model. This conversion follows a three-stage reasoning process inspired by the Chain-of-Thought (CoT) paradigm, as illustrated in Figure~\ref{fig:plan}. The process consists of the following steps:

\begin{itemize}[leftmargin=*, topsep=3pt, itemsep=0pt]
\item \textbf{Event and Timing Planning.} Given the input caption, the LLM first identifies a set of distinct audio events $\mathcal{E} = \{e_i\}_{i=1}^N$. For each event, it infers a corresponding set of timing spans $\mathcal{T}_i = \{(s_{i1}, t_{i1}), \dots\}$, where $s_{ik}$ and $t_{ik}$ are the start and end times in seconds. This multi-span representation is designed to handle events that occur multiple times.

\item \textbf{Speech Content Planning.} For any event identified as speech ($e_i \in \mathcal{E}_{\text{speech}}$), the LLM then infers a plausible utterance $c_i$ that fits the overall context. This step enriches the planned events with specific, intelligible speech content, resulting in a set of intermediate tuples $(e_i, \mathcal{T}_i, c_i)$.

\item \textbf{Prompt Recaption.} Finally, the LLM serializes the extracted information into the final structured prompt ($y_s$). This process starts with the original caption ($y_c$) and appends a specially formatted string for each planned event, which includes its name, associated time spans, and any inferred speech content.
\end{itemize}

\begin{table*}[h]
\centering
\caption{Examples of LLM-based planning results, converting natural language inputs (caption and speech text) into structured prompts.}
\label{tab:planning_examples}
\begin{tabularx}{\linewidth}{>{\raggedright\arraybackslash}X >{\raggedright\arraybackslash}X}
\toprule
\textbf{Input} & \textbf{Generated Structured Prompt (Output)} \\
\midrule

\textbf{Caption:} She is talking in the park. \newline
\textbf{Text:} "Good morning! How are you feeling today?"
& 
She is talking in the park. @\{park ambient sounds. \& <0.00, 10.00>\}@\{Female speech, woman speaking. \& <1.50, 6.00> "Good morning! How are you feeling today?"\} 
\\
\midrule

\textbf{Caption:} A child yelling as a young boy talks during several slaps on a hard surface \newline
\textbf{Text:} "Say yeah, baby. Say yeah, baby. Are you over tired?"
& 
A child yelling as a young boy talks during several slaps on a hard surface. @\{Young boy speaking \& <1.50,8.00> "Say yeah, baby. Say yeah, baby. Are you over tired?"\} @\{Child yelling \& <2.00,6.00>\} @\{slaps on a hard surface \& <2.50,3.00> <5.00,5.50>\} 
\\
\midrule

\textbf{Caption:} A female speaking with some rustling followed by another female speaking \newline
\textbf{Text:} "The IT services at the King's University College are proud to announce that we have launched"
& 
A female speaking with some rustling followed by another female speaking. @\{Female speech, woman speaking \& <0.50,6.00> "The IT services at the King's University College are proud to announce that"\} @\{rustling \& <1.00,5.00>\} @\{Female speech, woman speaking \& <6.50,8.00> "we have launched"\} 
\\
\midrule

\textbf{Caption:} A duck quacks followed by a man talking while birds chirp in the distance \newline
\textbf{Text:} "Mama Mama snow mama come over here, baby"
& 
A duck quacks followed by a man talking while birds chirp in the distance. @\{duck quack \& <0.50,1.50>\} @\{Man speaking \& <2.00,7.50> "Mama Mama snow mama come over here, baby"\} @\{birds chirping in the distance \& <2.50,4.00> <5.50,7.00>\} 
\\
\midrule

\textbf{Caption:} Two men speaking with loud insects buzzing \newline
\textbf{Text:} "I've got gloves covered in mid repellent. Still fishing."
& 
Two men speaking with loud insects buzzing. @\{Man speaking \& <1.00,4.50> "I've got gloves covered in mid repellent."\} @\{Man speaking \& <5.00,6.50> "Still fishing."\} @\{loud insects buzzing \& <0.00,10.00>\} 
\\
\midrule

\textbf{Caption:} A man speaking as a stream of water splashes and flows while music faintly plays in the distance \newline
\textbf{Text:} "in the amateur show tonight then tomorrow on Saturday the broadcasters and the other amateur cast will be going out hope to do well there get some good footage hope you enjoy"
& 
A man speaking as a stream of water splashes and flows while music faintly plays in the distance. @\{Man speaking \& <0.50,9.50> "in the amateur show tonight then tomorrow on Saturday the broadcasters and the other amateur cast will be going out hope to do well there get some good footage hope you enjoy"\} @\{water splashing and flowing \& <0.00,10.00>\} @\{faint music in the distance \& <0.00,10.00>\} 
\\
\midrule

\textbf{Caption:} People are giggling, and a man speaks \newline
\textbf{Text:} (None)
& 
People are giggling, and a man speaks. @\{people giggling \& <1.00,5.00>\} @\{Man speaking \& <2.50,4.50> "What's so funny?"\} 
\\
\midrule

\textbf{Caption:} (None) \newline
\textbf{Text:} "Some people talk about fucking the heads, but the way I do it, I just put my finger down there and pull it out."
& 
A person is giving instructions or explaining a procedure. @\{Man speaking \& <1.00,9.00> "Some people talk about fucking the heads, but the way I do it, I just put my finger down there and pull it out."\} 
\\

\bottomrule
\end{tabularx}
\end{table*}

\subsection{Planning Results for AC-Filtered}
\label{sec:plan-ac}

To qualitatively assess the effectiveness of our LLM-based prompt planner, Table~\ref{tab:planning_examples} presents several example results on samples from the AC-Filtered dataset. The table illustrates the planner's capability to parse complex, free-form captions (with or without associated speech text) and convert them into the precise, machine-readable structured prompts that our framework requires. This planning process is particularly crucial for enabling complex, multi-speaker scenarios. For instance, the planner can generate prompts that assign different utterances to distinct speakers at specified times. This capability stands in sharp contrast to speech-oriented models like VoiceLDM, which, even when given a descriptive prompt about a conversation, can only render the entire speech content as a single utterance from one voice. This ability to plan and generate true dialogues is a key advantage of our approach for creating realistic acoustic scenes.

\section{Speech Transcription via ALM}
\label{sec:alm}

\begin{figure*}[h]
\centerline{\includegraphics[width=17cm]{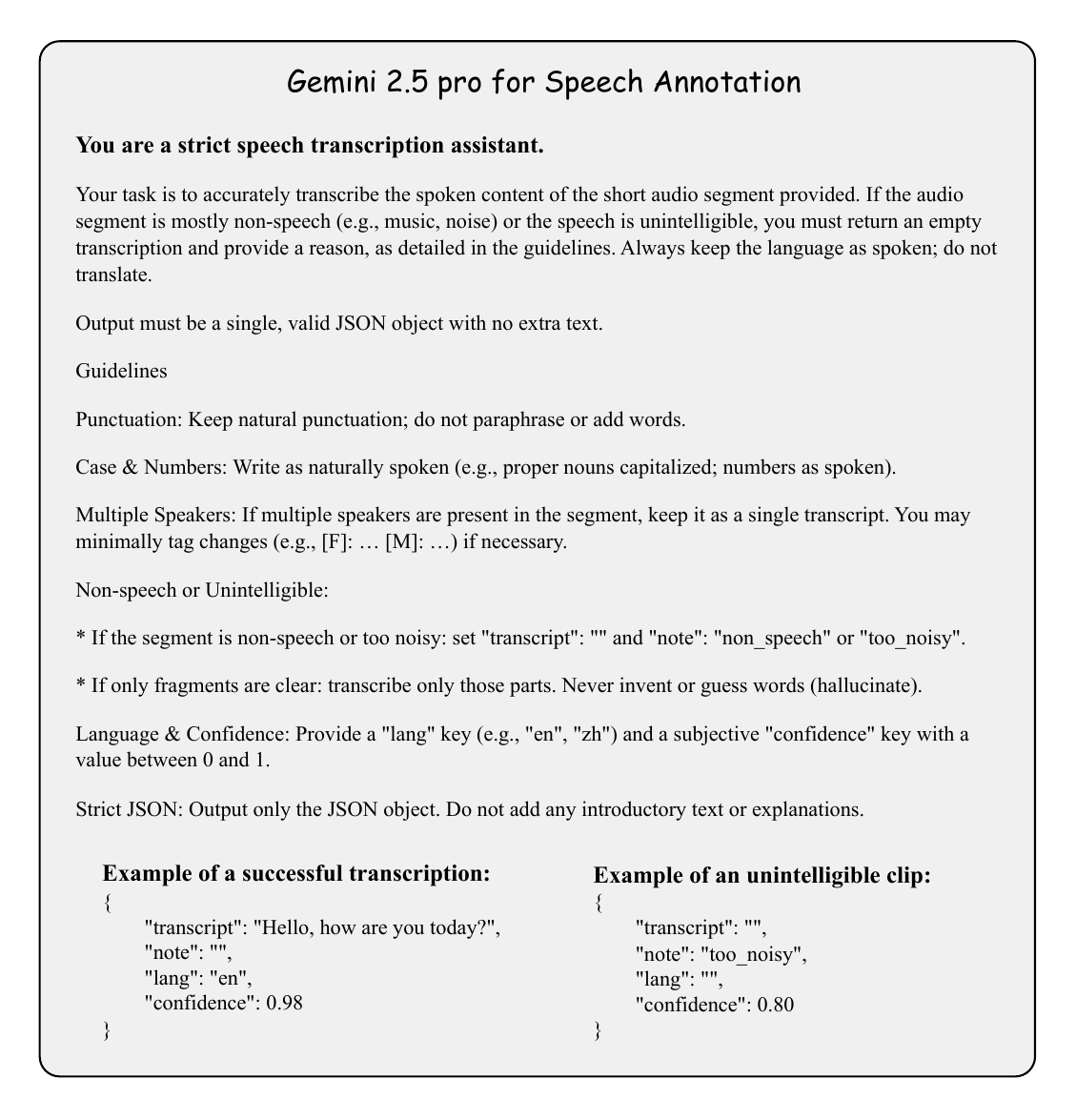}}
\caption{Gemini 2.5 pro for Speech Annotation.}
\label{fig:gemini_prompt}
\end{figure*}

To generate textual transcriptions for our segmented speech events, we utilize the Gemini 2.5 Pro model. Each clean audio segment is provided as direct input. We designed a prompt that serves a dual function: it instructs the model to accurately transcribe the spoken content while simultaneously acting as a quality filter. Specifically, the prompt directs the model to return an empty string if the speech in an audio segment is unintelligible or heavily obscured by noise, thereby automatically discarding low-quality samples. This process ensures that only clear, valid audio segments are converted into high-quality audio-text pairs. The full prompt used for this task is illustrated in Figure~\ref{fig:gemini_prompt}.

\end{document}